# A morphogenetic crop model for sugar-beet (Beta vulgaris L.). [‡]

Sébastien Lemaire[1,2,3], Fabienne Maupas[1]*, Paul-Henry Cournède[4,5], Philippe de Reffye[2,5]
(1 Institut Technique de la Betterave, Paris, 75008 France)
(2 CIRAD, UMR AMAP, Montpellier, 34000 France)
(3 AgroParisTech, UMR EGC, Grignon, 78000 France)
(4 Ecole Centrale Paris, Laboratoire MAS, Châtenay Malabry, 92295, France)
(5 INRIA Saclay- Île de France, EPI Digiplante, 91890 Orsay, France)

**Abstract:** This paper is the instructions for the proceeding of the International Symposium on Crop. Sugar beet crop models have rarely taken into account the morphogenetic process generating plant architecture despite the fact that plant architectural plasticity plays a key role during growth, especially under stress conditions. The objective of this paper is to develop this approach by applying the GreenLab model of plant growth to sugar beet and to study the potential advantages for applicative purposes.
Experiments were conducted with husbandry practices in 2006. The study of sugar beet development, mostly phytomer appearance, organ expansion and leaf senescence, allowed us to define a morphogenetic model of sugar beet growth based on GreenLab. It simulates organogenesis, biomass production and biomass partitioning. The functional parameters controlling source-sink relationships during plant growth were estimated from organ and compartment dry masses, measured at seven different times, for samples of plants. The fitting results are good, which shows that the introduced framework is adapted to analyse source-sink dynamics and shoot-root allocation throughout the season. However, this approach still needs to be fully validated, particularly among seasons.

## 1 Introduction

For healthy and unstressed crops of sugar beet, the total amount of dry matter is proportional to the amount of radiation intercepted by the canopy during the growth (Jaggard and Qi 2006). The leaf area controls the interception of radiation and its expansion is particularly important until full leaf cover is reached. In sugar beet, a leaf area index of about 3.0 is needed for maximal interception (Malnou et al. 2008). Therefore, any factor restricting the speed of leaf surface expansion directly reduces the final production. Increase in leaf area index depends on the rate at which new leaves appear and expand, on their final sizes and on how long they are retained by plants. All these factors are strongly influenced by the environment (climate, irrigation, fertilization) (Milford et al. 1985a). In stress situations, there is a strong interaction between plant growth and architectural development (Werker et al 1999) and classical crop models are not able to predict accurately root biomass and sugar content. In a context of sustainable agriculture and low input crop management, it is important to better understand plant physiology under stressed conditions and a model that takes into account plant morphogenesis could be a useful tool. In this prospect, this paper aims at introducing a morphogenetic model of sugar beet growth and study how this model can help understanding the shoot-root interaction during plant growth.

---

[‡] *S. Lemaire, F. Maupas, P.-H. Cournède, and P. de Reffye. A morphogenetic crop model for sugar-beet (beta vulgaris l.). In International Symposium on Crop Modeling and Decision Support: ISCMDS 2008, April 19-22, 2008, Nanjing, China, 2008.*

Functional-Structural plant growth models combine the description of the architectural development of plants and of the ecophysiological processes governing resource acquisition and repartition. We refer to Sievänen et al (2000) for the presentation of general concepts and reviews or de Reffye et al (2008) for the presentation of the most recent progresses.

GreenLab is such type of model. It takes its origin in the AMAP architectural models (de Reffye, 1988) but its ecophysiological concepts are inspired from those classically used in process-based models (Monteith 1977; De Wit 1978; Howell and Musick 1985; Marcelis et al 1998 or Qi et al 2005 for sugar beet), except that the dynamics of source-sink interaction is described at the level of organs according to their rhythm of appearance, see de Reffye and Hu (2003). The model does not claim to be fully mechanistic with regards to physiological and biophysical processes and fluxes involved in plant growth but a particular care is taken to follow empirically the dynamics of the carbohydrate budget, production and allocation, see (Yan et al 2004).

Its mathematical formulation as a discrete dynamical system (Cournède et al 2006) and the relative low number of parameters makes it suitable to identification from experimental data, in order to test its predictive capability. The model parameters have already been estimated for several plant species: sunflower (Guo et al 2003), maize (Guo et al 2006; Ma et al 2008), cucumber (Mathieu et al 2008), tomato (Dong et al 2008) among others. Model predictive ability was studied in details in (Ma et al 2007).

For this reason, it seemed interesting to use this model in order to describe the dynamics of source sink interaction during the growth of sugar beet, more precisely the balance between the vegetative part and the root system.

This paper begins with a quick presentation of the main concepts underlying the GreenLab model and of its adaptation to the sugar beet plant. The experimental protocol carried out in 2006 in order to collect the experimental data necessary for the model calibration is then described. In the second part of the article, we present the main results of the study. With the data collected at different stages during plant growth, the parameters of the GreenLab model can be estimated, which allows quantifying precisely the source-sink dynamics. Finally, these results are discussed in order to open new research perspectives with the objective to develop new tools for yield prediction and optimization.

## 2 Materials and methods

In this section, we briefly recall the basic concepts underlying the GreenLab model and the adapted experimental protocol for its parametric identification. The experimental data collected in 2006 are then given. A more detailed presentation of the model can be found in de Reffye and Hu (2003) and Yan et al (2004).

**A source-sink model with a common pool of reserves**
The main hypothesis to compute the functional growth is that the biomass produced by each leaf is stored in a common pool of reserves and redistributed among all organs according to their sink strengths. The initial seed and the leaves are sources. Petioles, blades and the root system are sinks.

The time unit to compute the ecophysiological functioning (resource acquisition and allocation) is chosen to coincide with the time unit of the morphogenetic sequence based on phytomer appearance. This time unit is called growth cycle (GC) and is thus classically given in thermal time by the phyllochron, that is to say the sum of degree-days necessary for a new phytomer to develop, cf. Dale and Milthorpe (1983).

Therefore, the individual plant is described as a discrete dynamical system. At growth cycle n, the empirical equation of neat dry matter production $Q_n$ is given by

$$Q_n = PAR_n \, \mu \, S^p \left( 1 - \exp\left( -k_B \frac{S_n}{S^p} \right) \right) \quad (1)$$

- $PAR_n$ denotes at cycle $n$ the incident photosynthetically active radiation. It is assumed to equal 0.48 times the global incident radiation (RG), *cf.* Gallagher (1978), Varlet-Grancher (1989).
- $\mu$ is an empirical coefficient related to the Radiation Use Efficiency,
- $S^p$ is an empirical coefficient corresponding to a characteristic surface (related to the two-dimensional projection of space potentially occupied by the plant onto the x-y plane)
- $S_n$ is the total leaf surface area of the plant at cycle $n$
- $k_B$ is the Beer-Lambert extinction coefficient.

In our equation, the ratio $\frac{S_n}{S^p}$ can be seen as a "local Leaf Area Index", see Cournède et al (2008).

At every growth cycle, the biomass thus produced is allocated to organs individually according to their relative demands called sink strengths. The sink strength of an organ depends on its type (blade, petiole and root in sugar beet) and varies from its initiation to maturity which corresponds to the end of its expansion. For an organ of type o (o=b,p,r for blade, petiole and root respectively), the sink variation $P_o$ is given classically in GreenLab (*cf.* de Reffye and Hu 2003) as a function of its age $j$ (in terms of growth cycles) as follows:

$$P_o(j) = p_o f_{a_o, b_o}(j) \quad \text{for} \quad 0 \leq j < T_o \quad \text{and} \quad P_o(j) = 0 \text{ otherwise.} \quad (2)$$

with $f_{a_o, b_o}$ a normalized beta distribution and $T_o$ the time (in growth cycles) necessary for the organ to reach its maximal size from its initiation and named "expansion time". However, in sugar beet, the expansion time widely varies from one blade to another, from one petiole to another. As a consequence, it is necessary to determine experimentally the expansion times of each blade and each petiole according to their rank $k$: $T_{b,k}$ and $T_{p,k}$. Moreover, we also had to adapt the sink variation function in order to take this phenomenon into account. Several tests and trials led us to choose for $P_{o,k}(j)$, the sink variation of an organ of type $o$, of rank $k$ and of chronological age $j$:

$$P_{o,k}(j) = p_o f_{a_o,b_o}\left(\frac{T_o}{T_{o,k}} j\right) \tag{3}$$

and

$$f_{a_o,b_o}(j) = \frac{1}{M_o}\left(\frac{j+0.5}{T_o}\right)^{a_o-1}\left(1-\frac{j+0.5}{T_o}\right)^{b_o-1} \tag{4}$$

with $M_o$ chosen such that: $\sup_j f_{a_o,b_o}(j) = 1$ and $T_o = \max_k T_{o,k}$.

A specific change in biomass allocation at canopy closing led us to consider a variable petiole sink:

$$P_{p,k}(j) = (p_p + q_p I_k) f_{a_o,b_o}\left(\frac{T_o}{T_{o,k}} j\right) \tag{5}$$

where $I_k$ denotes a competition index at growth cycle k and is given by:

$$I_k = 1 - \frac{S^p\left(1-\exp\left(-k_B \frac{S_k}{S^p}\right)\right)}{k_B S_k} \tag{6}$$

At each growth cycle n of its expansion period, an organ of age $i$ receives a biomass increment $\Delta q_o(n,i)$:

$$\Delta q_o(n,i) = P_o(i)\frac{Q_n}{D_n} \tag{7}$$

and its accumulated biomass $q_o(n,i)$ is thus given by the sum of all these increments since its appearance:

$$q_o(n,i) = \sum_{j=0}^{i} \Delta q_o(n-i+j, j) \tag{8}$$

If we assume a constant specific blade mass (SBM), the surface area of a given leaf is directly deduced by dividing by SBM the accumulated biomass of its blade. The total green leaf area $S_n$ used in the production equation is the sum of the surface areas of all the non senescent leaves. It implies determining for all phytomers the leaf life-span, that is to say the number of growth cycles between appearance and senescence. $T_{s,k}$ will denote the life-span (in growth cycles) of the leaf borne by the phytomer of rank $k$. If the phyllochron, expansion duration ($T_{b,k}$, $T_{p,k}$, $T_r$), life-span ($T_{s,k}$) and specific blade mass can be observed experimentally, it is not the case for the parameters: $\mu$, $S^p$, $(p_o, a_o, b_o)_{o=r,p,b}$. They will be estimated from experimental data by model inversion, as detailed by Guo et al (2006).

**Field experiments**
Field experiments were conducted in 2006 to investigate the sugar beet development of leaves and the growth of organs (root, blades and petioles). The experiments were carried out in 2006 in the Beauce plain near Pithiviers, France N48°10'12', E2°15'7'. A commercial sugar beet variety, Radar was sown on March 20 in husbandry conditions. The most uniform sections within a large sugar beet field were selected for the trials after plant emergence. This emergence stage (corresponding to the date when 80% of the final population is reached) occurred on April 8, corresponding to 150°Cdays (base temperature: 0°C) after sowing. The final population was estimated to be 9.6 plants per $m^2$. Daily mean values of air temperature (°C), solar radiation (MJ.$m^{-2}$), relative humidity (%) as well as total daily rainfall (mm) were obtained from French meteorological advisory services (Météo France) 5 km away from the experimental site. Thermal time was calculated by daily integration of air temperature (base temperature: 0°C) cumulated from emergence. The final harvest was carried out on October 3. The plants were given adequate water and fertilisers and were kept from pests and diseases.

**Development measurements**
Leaf development (appearance, expansion and senescence) was measured weekly non-destructively on the same groups of seven representative and adjacent plants. Coloured rings were placed around the petioles of the $1^{st}$, $5^{th}$, $15^{th}$ and $20^{th}$ leaves as markers.

*Leaf appearance and phyllochron:* the phyllochron is defined as the thermal time interval that separates the emergence of successive leaves, each corresponding to a phytomer (the two cotyledons forming the first phytomer). We consider a leaf as emerged when its length is above 10mm.

*Expansion:* blade lengths and widths as well as petiole lengths were measured to determine expansion kinetics. The curve giving the product of blade length by blade width as a function of the thermal time was fitted with a logistic equation with three parameters.

$$f(x) = \frac{B}{1 + \exp\left(\frac{b-x}{a}\right)}$$

with $x$ the thermal time from emergence.

Let $T_{i,k}$ be the thermal time corresponding to the appearance of phytomer $k$. Since, the parameter b corresponds to the inflexion point and since the curve is symmetrical, the expansion times for blades and petioles is: $T_{b/p,k} = 2(b - T_{i,k})$.

*Senescence:* a leaf was supposed senescent when its entire surface had yellowed. We thus determined $T_{s,k}$ for all $k$.

**Biomass measurements:**
Biomass measurements were carried out at seven different stages during the growing period: May 12 (423°Cdays), May 17 (499°Cdays), May 24 (597°Cdays), June 15 (989°Cdays), July 11 (1538°Cdays), August 8 (2178°Cdays) and October 3 (3168°Cdays). At each date, seven individual plants were selected (randomly) in a group of fifteen adjacent plants and the dry mass of every individual organ (blades, petioles and root storage) was measured. Dry matter was obtained by drying for 48h at 75°C. Every leaf was digitalized in order to estimate length and width of its blade and petiole, and blade surface area. In each group of fifteen, the eight other plants were measured at the level of organ compartments: total dry mass of blades, that of petioles and that of root. The final stage of measurements corresponds to harvest. For this date, the seven plants fully described (at the level of organs) are those for which the development scheme was established with non-destructive measurements.

These biomass measurements are the experimental data from which the model parameters are estimated, *cf.* Guo et al (2006) for the details of the calibration procedure.

## 3 Results and discussion

**Analysis of experimental data of sugar beet development and growth**
*Dry matter production and allocation*
The total dry matter per square meter is shown to be proportional to the amount of photosynthetically active radiation intercepted by the crop (PARi), *cf.* Fig 1, with at growth cycle *n*:

$$PARi_n = 0.48\, RG\left(1 - \exp(-k\, LAI_n)\right)$$

*RG* is the amount of global solar radiation.

$LAI_n$, the leaf area index at cycle n, is obtained by multiplying the average of the observed leaf area per plant by the crop density (9.6 plants per m$^2$). The ratio of dry matter production to intercepted PAR is defined as the Radiation Use Efficiency (RUE). It is widely used in crop models since Monteith (1977). For our experiments, the radiation use efficiency is found to be 3.36 g.MJ$^{-1}$. It is comparable to Damay's result (1993) who found 3 to 3.8 g.MJ$^{-1}$. Jaggard (2006) obtained an equivalent RUE of 3.66.MJ$^{-1}$.

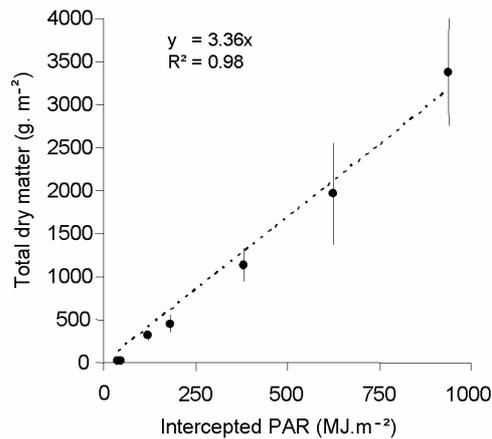

Fig. 1 Dry matter production against accumulated intercepted PAR

Fig. 2 a) shows the observed total dry matter accumulated in the three compartments (roots, blades, petioles). After 750 to 800°Cdays, most assimilates are allocated to the storage root. In Fig. 2 b), the ratio of blade dry matter to petiole dry matter against thermal time since emergence illustrates the partition between exchange organs (blades) and structural organs (petioles). At the first stages of growth, with little competition for light, exchange surface (leaf surface area) is privileged. With increasing intra- and inter- plant competition, allocation to petioles relatively increases in order to face competition effects.

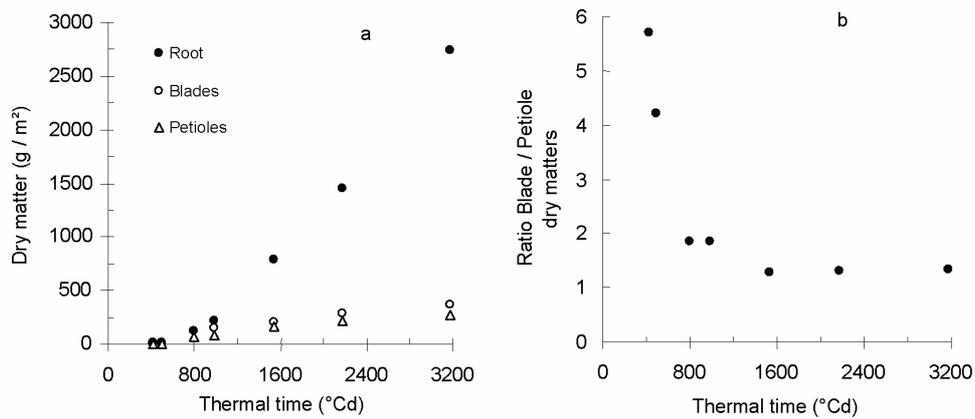

Fig. 2 Dry matter repartition between the different compartments against thermal time since emergence. (a) Accumulated dry mass of roots (closed dots), blades (open dots) and petioles (open triangles). (b) Ratio of accumulated blade dry mass to petiole dry mass.

*Phytomer emission and phyllochron*
Growth pattern of successive leaves depend on both ontogeny and temperature. The effect of temperature can be integrated by expressing leaf appearance and expansion as linear functions of thermal rather than chronological time (Milford 1985a). The first pair of leaves (cotyledons) unfolds from the apex together; subsequent leaves appear individually on a 5:13 phyllotaxis, as described by Stehlik, 1938 and Milford, 1985b. We thus consider that the pair of cotyledons forms the first phytomer and each subsequent leaf corresponds to a phytomer. For all the seven plants, plotting the number of phytomers against thermal time since emergence reveals a piecewise linear relationship, with two distinct intervals, *cf.* Fig. 3. The linear relation indicates that the temperature is the main determinant of the rate of leaf appearance and this rate is constant for each phase. The inverse of the slope is the phyllochron, *i.e.* the thermal time interval between the visual appearances of two successive phytomers (Milford *et* al 1985; Granier et al 1998).

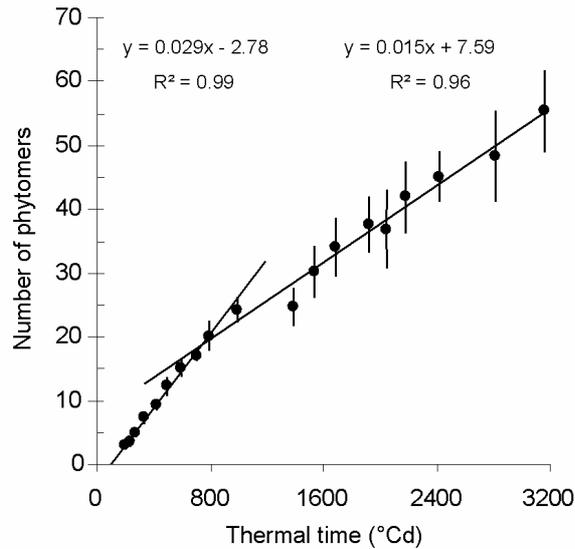

Fig. 3 Number of phytomers as a function of thermal time (base temperature: 0°C) since emergence. The dots correspond to the average number of phytomers for the 7 measured plants; error bars indicate standard deviations.

The regression slopes vary significantly for the seven plants, see Table 1.

Table 1: Phyllochrons of the two development phases for the extremes seven measured plants. The last line corresponds to the regression of the average number of phytomers at each measurement stage.

|  | First phase phyllochron (°Cdays) | $R^2$ | Second phase phyllochron (°Cdays) | $R^2$ |
|---|---|---|---|---|
| **Minimum** | 29.9 | 0.98 | 56.8 | 0.94 |
| **Maximum** | 38.5 | 0.99 | 78.1 | 0.95 |
| **Average plant** | **34.4** | **0.99** | **66.7** | **0.97** |

This change in phyllochron after approximately 20 leaves was already observed by Milford (1985a,b). He showed that the phyllochron of the first phase was constant among seasons and experimental treatments applied to the crop (sowing dates, N-content or plant density). Nevertheless, for these various treatments, the thermal duration of this early phase was variable, from 285 to 883°Cdays (thermal time from sowing with a base temperature of 1°Cday). In the second phase, the leaf appearance is slower (66.7°Cdays against 34.4°Cdays). Many hypotheses were suggested by Milford to explain the curve bending: base temperature that changes when the plant gets older, photoperiodic factor, trophic competition, that is to say the competition for assimilates between the developing storage root and vegetative organs. This competition may slow down the rate of leaf appearance. In our experiments, this rate starts decreasing at the beginning of the linear phase of root growth (Fig. 4.). This stage also corresponds to the observed canopy closing. Caneill (1994) also pointed out that there was important modification in assimilate partitioning between leaves and roots at canopy closing.

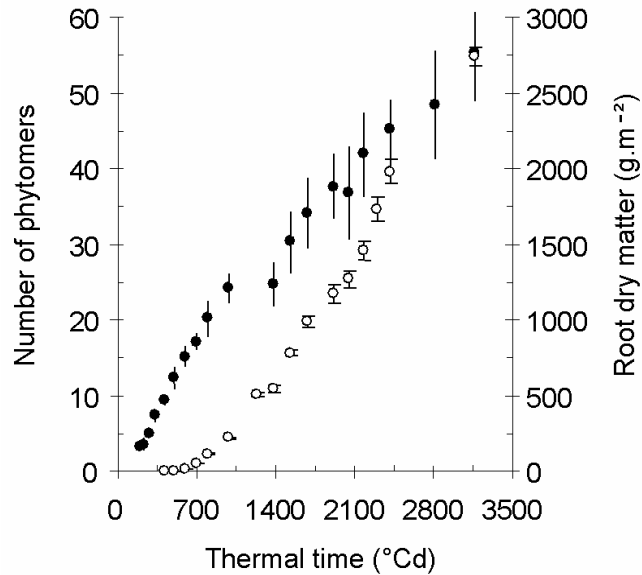

Fig. 4 Comparison of phytomer appearance (closed dots) and increase in root dry matter (open dots) as functions of thermal time since emergence. Arrows indicate the time of the observed canopy closing, which corresponds to the decrease of the rate of leaf emission and the beginning of the linear phase of root growth.

*Organ expansion and leaf senescence*
To control the dynamics of sources and sinks, it is necessary to determine the duration of the expansion phase for all organs (which says how long they are sinks) and the life-span of all leaves (which says how long they are sources). Fig. 5 compiles the thermal time since emergence of appearance, end of expansion and senescence for the leaves of all phytomers. There is again a change of incline, for both end of expansion and senescence, at canopy closing.

The expansion durations of petioles are really close to those of blades and are thus considered identical. In previous studies (unpublished data), the observation of root growth over two years gave us an expansion duration of 3900°Cdays for the root storage.

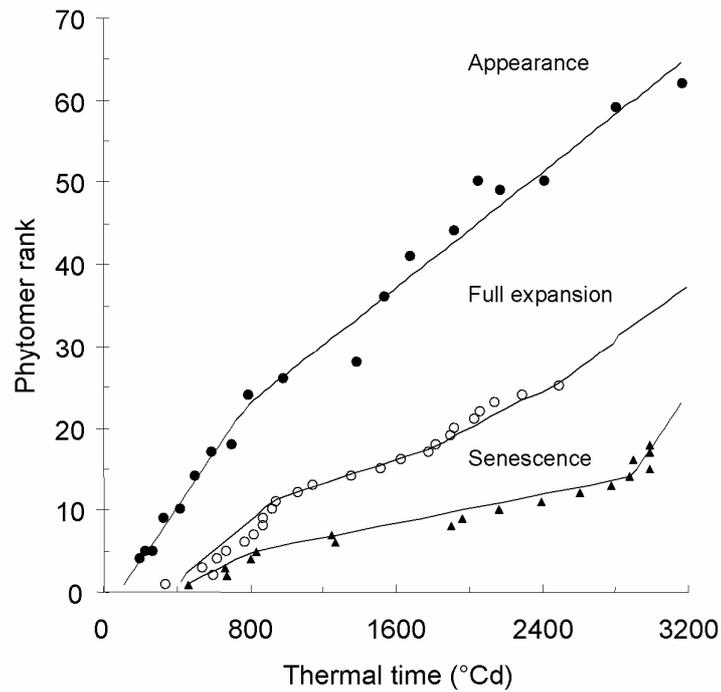

Fig. 5 Leaf development scheme

*Specific blade mass*
The specific blade mass is defined as the ratio between blade dry mass and blade surface area. Plotting one variable against the other for all measured data (different plants at different ages), *cf.* Fig. 6, shows that the linear relation is pretty good, even though the dispersion increases for bigger leaves.

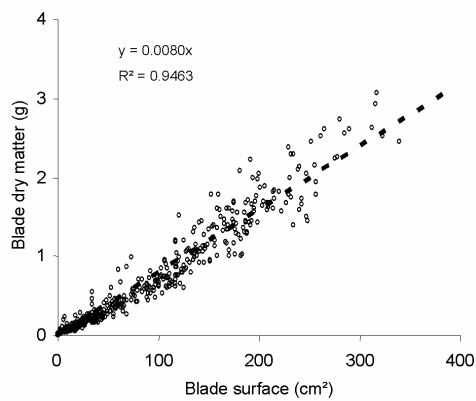

Fig. 6 Blade dry mass against blade surface for all measured data and results of the linear regression

The specific blade mass thus obtained is 0.0080g.cm$^{-2}$. It is useful in the GreenLab model to deduce the production of blade surface area from the accumulated biomass. Our results do not confirm a possible increase in specific blade mass when leaves get older, phenomenon underlined by Ann Clark and Loomis (1978).

**Identification of the morphogenetic model and source-sink dynamics**
The preliminary analysis of our experimental data helped us construct the morphogenetic model of growth based on GreenLab. First, the time step for model computation (also called Growth Cycle) is chosen as the smallest phyllochron observed. It allows that no phytomer emission is missed for any plant. Even after the increase of the phyllochron at canopy closing, the model time step remains the same but a new phytomer is not generated at every growth cycle in the model. A new phytomer appears or not according to the theory of discrete lines (Reveilles 1991), which ensures that at every growth cycle the number of phytomers is the best possible approximation of the continuous reality.

The expansion durations and life-spans for all organs according to their ranks which were obtained in thermal time are given their corresponding values in growth cycles. Some other parameters are measured (specific blade mass) or obtained in the literature (the Beer-Lambert extinction coefficient $k_B = 0.7$ according to Andrieu et al 1997). The environmental input $PAR_n$ is given by the cumulated photosynthetically active radiation at each growth cycle.

The other parameters are estimated from our experimental data. Note that since sink values are barycentric coefficients, one of these values must be fixed. For this reason, we impose that the blade sink strength is equal to 1. For the root sink strength, any important value would do since after some time the most important part of matter is allocated to the root.
Table 2 gives the results of the estimation procedure (achieved with Digiplante software, developed at Centrale Paris, *cf.* Cournède et al 2006).

Table 2: Parameters of the GreenLab model of sugar beet growth.

| Parameter | Description | Estimated (E) Measured (M) or Fixed (F) | Value | Unit |
|---|---|---|---|---|
| $\mu$ | Empirical coefficient related to the radiation use efficiency | E | 1.23 | g.MJ$^{-1}$ |
| $k_B$ | Beer-Lambert extinction coefficient | M | 0.7 | - |
| $S^p$ | Empirical coefficient corresponding to a characteristic surface | E | 0.021 | m² |
| SBM | Specific blade mass | M | 0.008 | g.cm² |
| $p_r$ | Root sink strength | F | 400 | - |
| $p_b$ | Blade sink strength | F | 1 (reference value) | - |
| $p_p$ | Petiole sink strength | E | 0.4916 | - |
| $q_p$ | Petiole sink correction | E | 0.3894 | - |
| $T_r$ | Root expansion duration | M | 130 | GC |
| $T_{b,k}$ | Blade expansion duration | M | Function of phytomer rank | GC |
| $T_{p,k}$ | Petiole expansion duration | M | Function of phytomer rank | GC |
| $T_{s,k}$ | Leaf life span | M | Function of phytomer rank | GC |
| $a_r$ | Parameter for beta law | E | 3.13 | - |
| $b_r$ | Parameter for beta law | E | 1.15 | - |
| $a_b$ | Parameter for beta law | E | 3.56 | - |
| $b_b$ | Parameter for beta law | E | 2.22 | - |
| $a_p$ | Parameter for beta law | E | 2.56 | - |
| $b_p$ | Parameter for beta law | E | 1.67 | - |

In Fig. 7 and 8, are shown the differences between the (averaged) experimental data and the model output obtained for the estimated parameters. The different stages of growth are satisfactorily reproduced at the level of organs, which proves that not only the global balances of biomass allocation are described but also the precise dynamics of source-sink interaction.

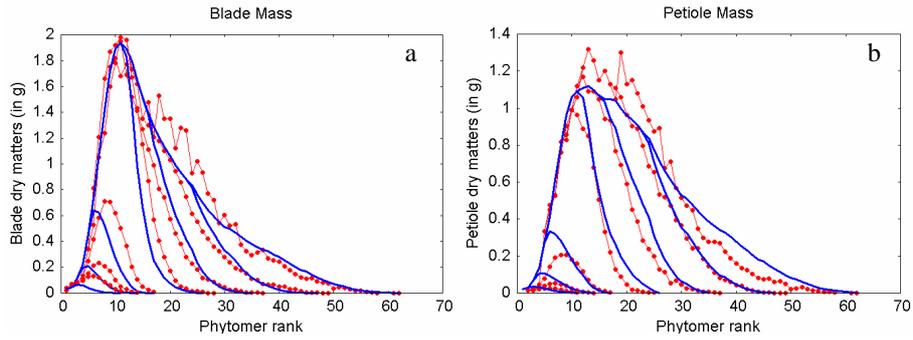

Fig. 7 Average plants at organ level: experimental data (open dots) and simulated data (lines) (a) Blade dry masses and (b) petiole dry masses according to phytomer ranks , at 423, 499, 597, 989, 1538, 2178 and 3168°Cdays after emergence.

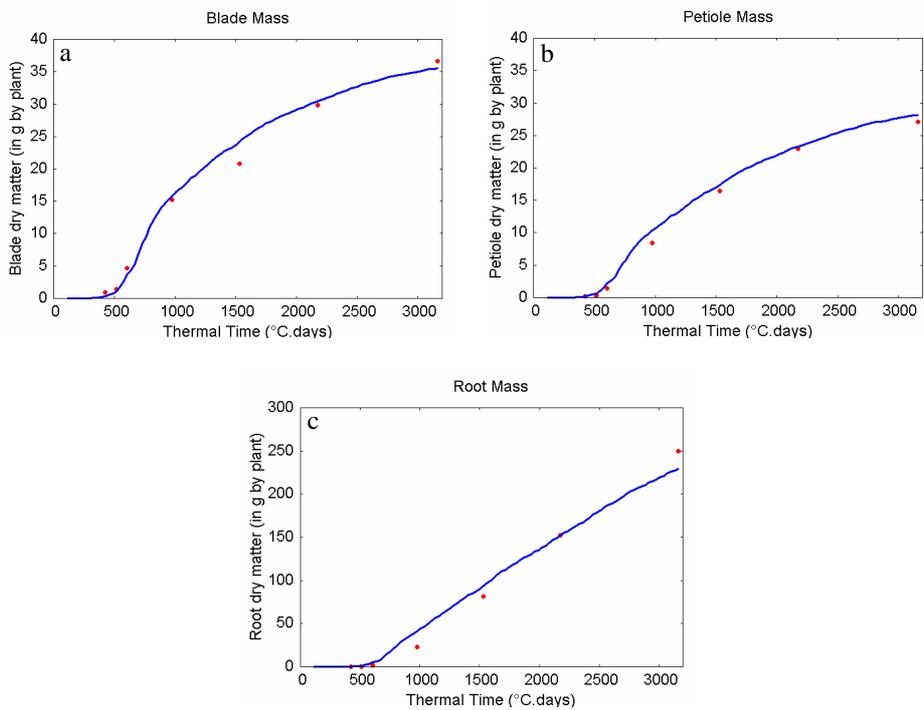

Fig. 8 Average plants at compartment level: experimental data (red dots) and simulated data (blues lines) (a) Blade compartment dry mass (b) Petiole compartment dry mass and (c) Root dry mass as functions of thermal time since emergence.

Fig. 9 shows the biomasses allocated to blades, petioles and root at every growth cycle. We clearly see the rapid increase of the allocation to the root compartment after 500 °Cdays since emergence with a maximal proportion of assimilates distributed to root storage at 700°Cdays after emergence, which corresponds to the maximal biomass production. This stage may be related to the linear phase of root growth and phyllochron change above-mentioned (*cf.* Fig.4). In Fig. 10 are shown the simulated proportions of biomass allocated to shoot and to root throughout the season. The observed tendencies are very classical and correspond to those given by the SUCROS model (Spitters et al 1989).

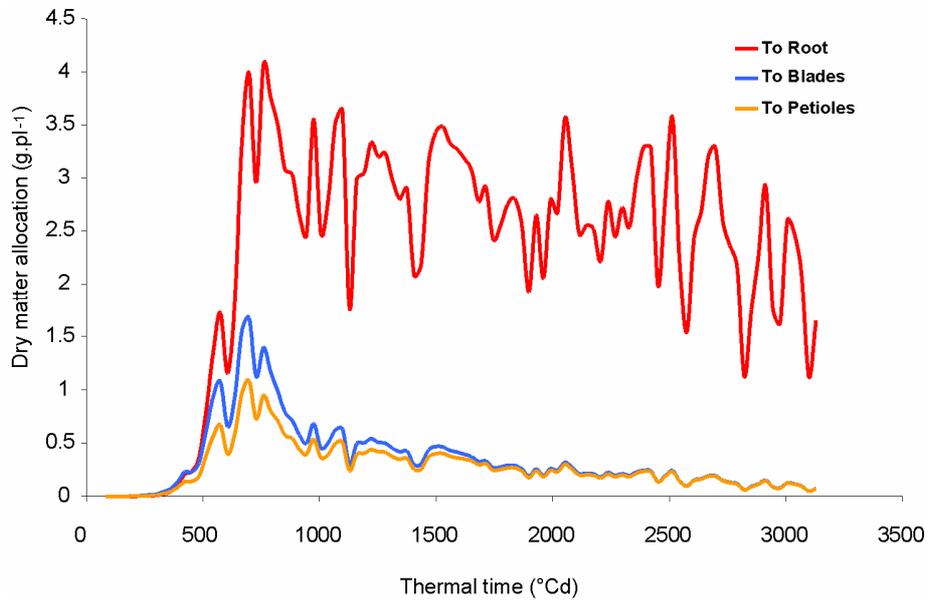

Fig. 9 Biomass allocation between blade, petiole and root compartments simulated by the GreenLab model

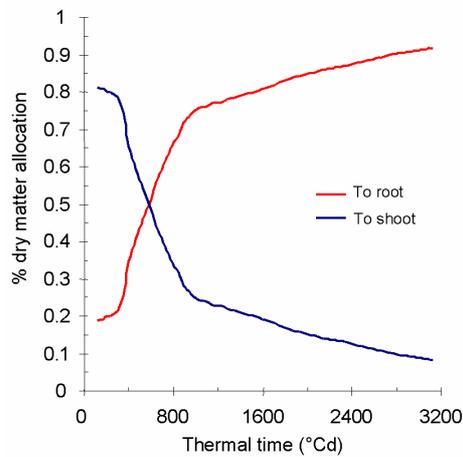

Fig. 10 Proportions of biomass allocated to shoot and to root throughout the season

# 4 Conclusion

The experimental study of sugar beet development, mostly phytomer appearance, organ expansion and leaf senescence, allowed us to define a morphogenetic model of sugar beet growth based on GreenLab. The functional parameters controlling source-sink relationships during plant growth were estimated from organ and compartment dry masses, measured at seven different times, for samples of plants. The fitting results are good and the dynamics of source-sink interactions is precisely described by the model throughout the season.

We believe such a simulation tool may be useful in a general context of input reduction in agricultural practices. The impacts of low input management need to be quantified. Estimation of losses in sugar yield resulting from stress is important for growers who need to make decisions for a rational use of inputs. If the relationship between root dry matter yield at harvest and light energy intercepted during growth is generally good, it shows limitations when plants are affected by stress, particularly water stress (Werker and Jaggard 1998). As foliage is the main determinant of root yield, it is important to precisely control leaf development and its interaction with root growth. Pests or water and nutrient limitations can cause sugar beet leaves to prematurely end their expansions or wilt. It entails a decrease of crop photosynthetic efficiency and, consequently, of root mass and sugar content.

Moreover, in sugar beet plants, there are numerous exchanges of carbohydrates between leaves and root storage, which are strongly modified when plant undergo stress. Therefore, it is important to better understand relationships between leaf development and growth throughout the growth season. Since the GreenLab model provides a dynamic description of matter allocation at the level of organs, changes in leaf development mechanistically impact the source-sink ratio (biomass production and allocation), and thus root growth. For this reasons, we believe the modelling framework introduced is a good candidate to help understand how stress affects the balance between sinks, more particularly the shoot-root allocation dynamics. Likewise, the morphogenetic model proposed may help determine characteristic physiological phases in sugar beet growth. Caneill (1994) proposed a description of different phases of sugar beet development depending on the priorities of biomass allocation among organs. He defined four particular stages according to the ratio between root dry matter and shoot dry matter. However, the thermal time necessary to reach these stages was not stable, in particular for the 3rd stage, which corresponds to the beginning of the linear root growth. The GreenLab model is potentially able to help determine these physiological phases, under different growing conditions including stress, density effects, sowing dates …

However, the preliminary results of this study simply prove that the introduced framework is adapted to analyse source-sink dynamics and shoot-root allocation. This approach still needs to be fully validated, particularly among seasons. More particularly, the leaf development scheme of several seasons and several stress conditions has to be studied. What are the system stabilities and what variables are strongly impacted? In such experimental conditions, how is the source-sink dynamics affected? New experiments were conducted in 2007 (including nitrogen stress and water stress) and in 2008 (4 different densities). The results are in the process of analysis.

## Acknowledgements

This work was supported by the French Sugar-beet Technical Institute (ITB). We thank A. de Silans, P. Houdmon, H. de Balathier, A. Nioche and M. Allart of ITB for technical assistance.